\documentclass[fleqn,usenatbib]{mnras}

\usepackage{newtxtext,newtxmath}

\usepackage[T1]{fontenc}

\DeclareRobustCommand{\VAN}[3]{#2}
\let\VANthebibliography\thebibliography
\def\thebibliography{\DeclareRobustCommand{\VAN}[3]{##3}\VANthebibliography}

\usepackage{graphicx}	
\usepackage{amsmath}	

\title[Boxy/peanut signatures in MW Miras]{Age-morphology dependence of the Milky Way boxy/peanut bulge seen in Mira variables}

\author[Semczuk et al.]{%
Marcin Semczuk,$^{\!\!1}$
Walter Dehnen,$^{\!\!2,1}$
Ralph Sch{\"o}nrich,$^{\!\!3}$ and
E. Athanassoula,$^{\!\!4}$
\smallskip
\\
$^1$ School for Physics and Astronomy, University of Leicester, University Road, LE1 7RH, UK \\
$^2$ Astronomisches Recheninstitut, Zentrum f{\"u}r Astronomie der Universit{\"a}t Heidelberg, M{\"o}nchhofstra\ss{}e 12-14, 69120, Heidelberg, Germany\\
$^3$ Mullard Space Science Laboratory, University College London, Holmbury St.~Mary, Dorking, Surrey, RH5 6NT, UK \\
$^4$ Aix Marseille Université, CNRS, CNES, LAM, Marseille, France
}

\date{Accepted XXX. Received YYY; in original form ZZZ}

\pubyear{2015}

\begin{document}
\label{firstpage}
\pagerange{\pageref{firstpage}--\pageref{lastpage}}
\maketitle

\begin{abstract}
We analyse the distribution of Mira variable stars in the central region of the Milky Way. We find that with increasing period, i.e. decreasing age, the Miras shift towards negative Galactic longitudes $\ell$. Comparing to a cosmological zoom simulation of a barred galaxy, we find that this shift with age can be explained by an age-morphology dependence of the boxy peanut/bulge. Owing to a combination of projection effects and the limitation of the range of Galactic longitudes, the near hump at $\ell>0$ is more truncated for younger populations, and the far hump at $\ell<0$ dominates the observed distributions.

\end{abstract}




\section{Introduction}
Boxy/peanut (B/P) bulges are frequently found in near edge-on barred galaxies \citep{Erwin2017}.  Near-IR images (\citealt{Weiland1994}; \citealt*{Ciambur2017}) and the split of the magnitude distribution of the red clump stars \citep{McWilliamZoccali2010, SaitoEtAl2011, Ness2012} suggest that the Milky Way also hosts a B/P bulge.

While these observational studies gives us a view of the B/P bulge that is comprised of stars from a wide range of ages, several authors (\citealt*{Athanassoula2017}; \citealt{Debattista2017}; \citealt{Fragkoudi2017}; \citealt{Buck2018}; \citealt{Fragkoudi2020}) showed that both in idealised $N$-body and hydrodynamical simulations the morphology of stellar populations in a B/P bar/bulge depends on age (or metallicity). \cite{Debattista2017} attributed this dependence to the kinematic `fractionation' of stellar orbits. In their simulations, the bar growth affects populations of different prior kinematics differently, i.e. populations with larger (smaller) prior scale-height will assume a more extended (less extended) vertical profile after bar formation. In \cite{Semczuk2022}, we added to this picture that inside-out formation of the disc prior to the bar formation induces a similar correlation of B/P extent vs. age.

The age-morphology dependence of bulges was confirmed in external galaxies \citep[nested structures,][]{Ciambur2016} and in the Milky Way, where the red clump split was observed in metal rich but not in metal poor stars \citep{Ness2012, Uttenthaler2012, Rojas-Arriagada2014, Babusiaux2016}. Aside from direct stellar parameter determination, the age-/metallicity-dependent prevalence of different types of variable stars can be used to trace different populations. In a recent paper \citep{Semczuk2022} we showed that RR Lyrae stars \citep[from the OGLE catalogue][]{Soszynski2014, Soszynski2019}, which represent a very old population (>10 Gyr), display signatures of a small peanut bulge, consistent with fractionation predictions.  

In this work, we further investigate the age-morphology dependence of the Milky Way bulge with the recent OGLE Mira stars catalogue \citep{Iwanek2022} by employing the Miras' period-age relation \citep{Trabucchi2022}. Section~\ref{sec:data} describes the data and section~\ref{sec:Auriga} briefly describes the simulations. In Section~\ref{sec:simulation} we discuss the predictions for the age-morphology dependence from the simulations. In Section~\ref{sec:Morph:data} we compare the age-/period-dependent distribution of OGLE Miras to these theoretical predictions and in Section~\ref{sec:discuss} we discuss our results in the context of previous studies of Mira stars in the Milky Way.

\section{Data Selection}
\label{sec:data}
In this study we make use of a recently published catalogue of OGLE Mira stars\footnote{\url{https://ogledb.astrouw.edu.pl/~ogle/OCVS/}} (\citealt{Iwanek2022}). We cross-matched the OGLE sample with 2MASS (\citealt{Skrutskie2006}) using 1 arcsec aperture to obtain the $JHK_s$ photometry. We cleaned the crossmatched sample taking only stars with colours $J-H>0.75$ by rejecting the clear outliers in $J-H$ vs $J$ colour-magnitude diagram. In order to avoid selection effects caused by OGLE footprints we restrict the data to the region $|\ell|<10\degr$ and $-5\degr<b<-2.8\degr$. This region also overlaps with our previous study of the B/P bulge in RR Lyrae stars \citep{Semczuk2022}. Similar to \cite{Grady2020}, we restrict the Miras' period range to 100-400 days, giving a final sample of 3446 stars.

\begin{figure*}
    \centering
    \includegraphics[width=17.85cm]{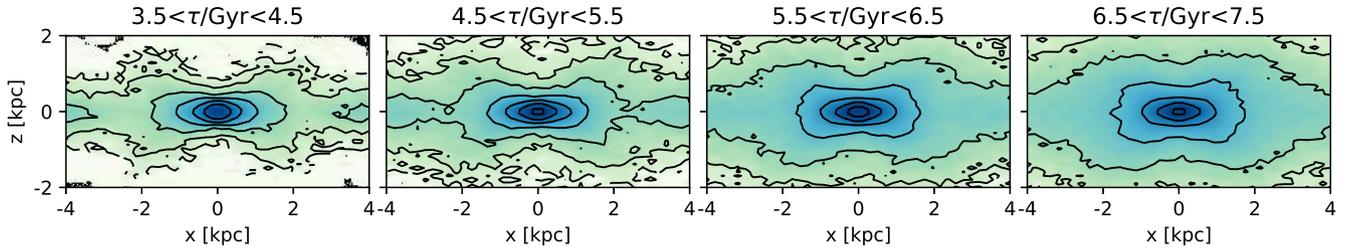}
    \vspace*{-6mm}
    \caption{Side-on projected density from a simulated barred galaxy (Auriga 6) for four stellar populations of increasing age $\tau$. To emphasize the morphology of the B/P bulge the particles were selected from $|y|<0.25$ kpc.
    \label{sim0}}
\end{figure*}
\begin{figure*}
    \centering
    \includegraphics[width=17.85cm]{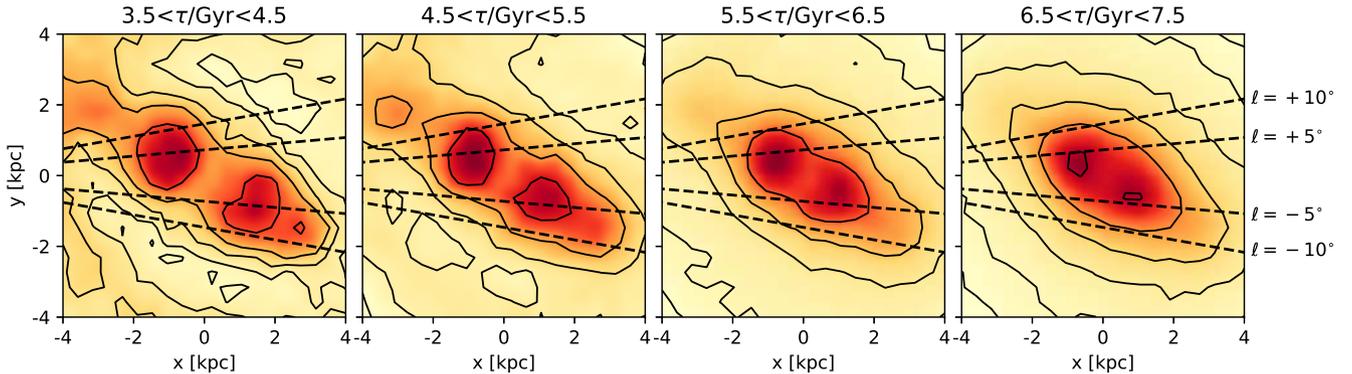}
    \vspace*{-6mm}
    \caption{Face-on surface density distributions of the Auriga 6 simulation for four different stellar populations selected by their age $\tau$. Stellar particles for these panels were selected from the range $2.8\degr<|b|<5\degr$ in order to mimic the selection criteria of the OGLE sample. Note. however, that for simulations both positive and negative $b $ are stacked to increase resolution, while for OGLE data we only have the negative side). Dashed lines outline cuts in Galactic longitude $\ell$ as indicated (calculated assuming the Solar position $x=-8.3$ kpc).}
    \label{sim1}
\end{figure*}

\section{The simulations}
\label{sec:Auriga}
In this paper we use publicly available snapshot at $z=0$ of the Auriga 6 simulation from the Auriga suite of cosmological zoom-in simulations of Milky-Way analogues \citep{Grand2017}. Auriga is a suite of 30 magneto-hydrodynamical runs done with the code \textsc{arepo} \citep{Springel2010}. The selected galaxy has a virial halo mass of $1.04\times10^{12}\;\mathrm{M}_{\odot}$, disc mass of $3.32\times10^{10}\;\mathrm{M}_{\odot}$ and disc scale length of $5.95$ kpc. The disc of the simulated galaxy also hosts a bar with length $<5$ kpc and an associated B/P bulge.
\section{Age-morphology dependence of the boxy/peanut bulge}
\label{sec:simulation}

Several authors \citep{Athanassoula2017, Debattista2017, Fragkoudi2017, Buck2018, Fragkoudi2020, Semczuk2022} have shown that the morphology of the B/P bulge differs between different stellar populations in $N$-body or hydrodynamical simulations. 

In \cite{Semczuk2022} we used growing N-body disc simulations to analyse how the age-morphology dependence of the peanut bulge manifests itself in samples covering windows of height and galactic latitude above/below the galactic disc. The cosmological zoom simulation examined here gives consistent results, which we will use as a qualitative example of how this age dependence is expected to imprint on the current observations.  

Figure~\ref{sim0} shows cross sections of side-on (along the bar major axis) stellar density distributions for four different bins in stellar age $\tau$. Similarly to simulations listed above, the younger populations have a more pronounced/higher constrast X shape in the outer parts, while in the inner parts the opposite is the case: their distribution is smoother/more oval compared to the older populations.

With fixed cuts in Galactic latitude the projected densities thus show a significant difference between populations in the separation of the peanut humps. Figure~\ref{sim1} shows clearly that the two over-densities are most separated for youngest stars (left) and come closer to each other for older populations (right). We note aside that typically the inclined sample geometry in altitude vs. distance plays a lesser role. More important is the different relative inclination of the sightlines to the orientation of the bar when projecting in Galactic longitude: here, the far hump dominates the on-sky number density for younger populations. Depending on limitation, the close hump (at positive $y$) gets more and more truncated by the longitude cut. For the oldest populations the distribution is rounder/less elongated and more of both humps fall into the longitude range \citep[as in the case of RR Lyrae stars in][]{Semczuk2022}. The left panels of Figures~\ref{front} and~\ref{hist} show this in the on-sky distribution and a histogram, respectively, for the same populations plotted in Figure~\ref{sim1}. Histograms show that the count of stars at negative longitudes gets higher with smaller ages, with respect to the peaks of the distributions.  

\section{Morphology of the OGLE Miras distribution }
\label{sec:Morph:data}

\begin{figure*}
    \centering
    \includegraphics[width= .98\columnwidth]{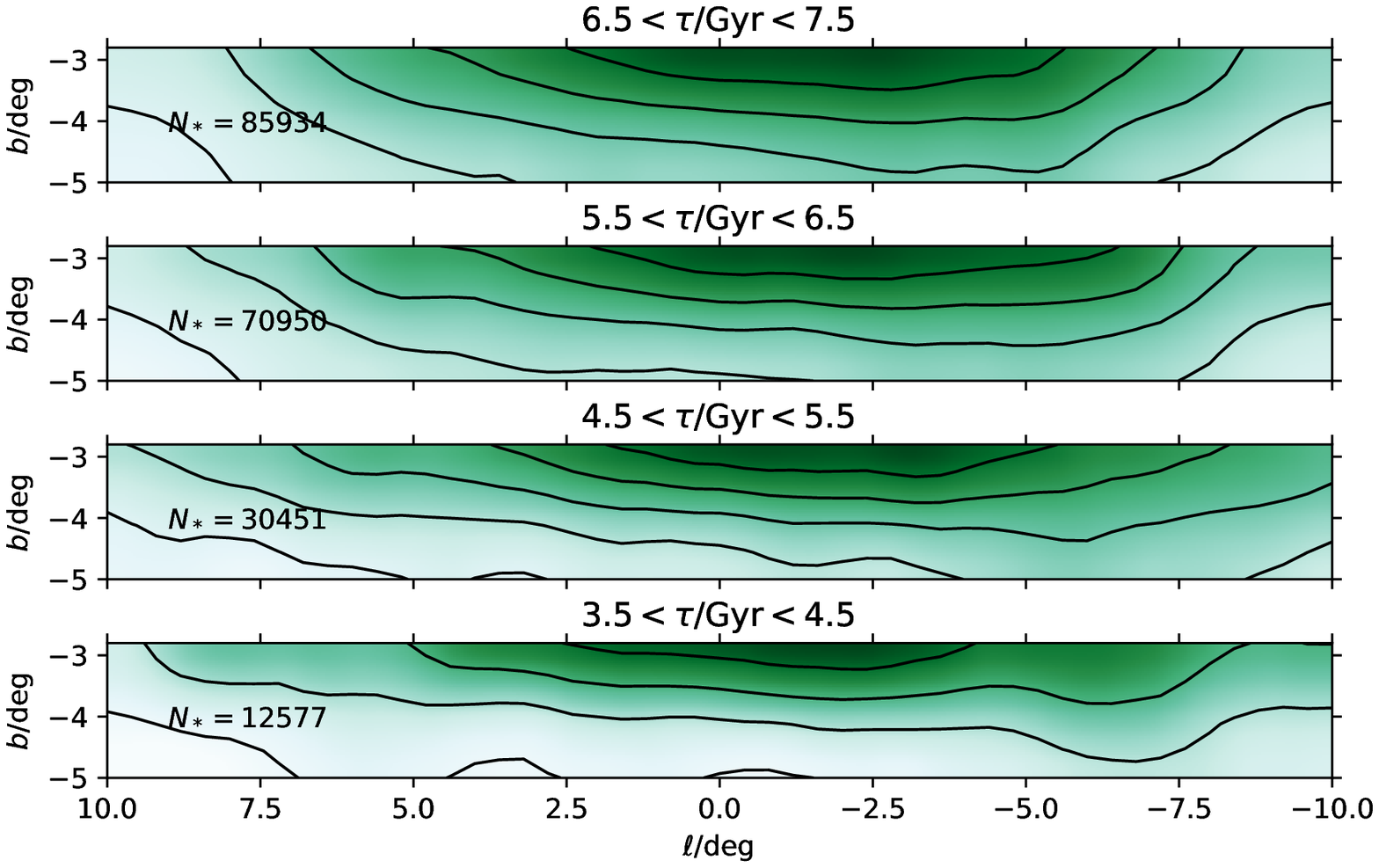}\hfil
    \includegraphics[width=.98\columnwidth]{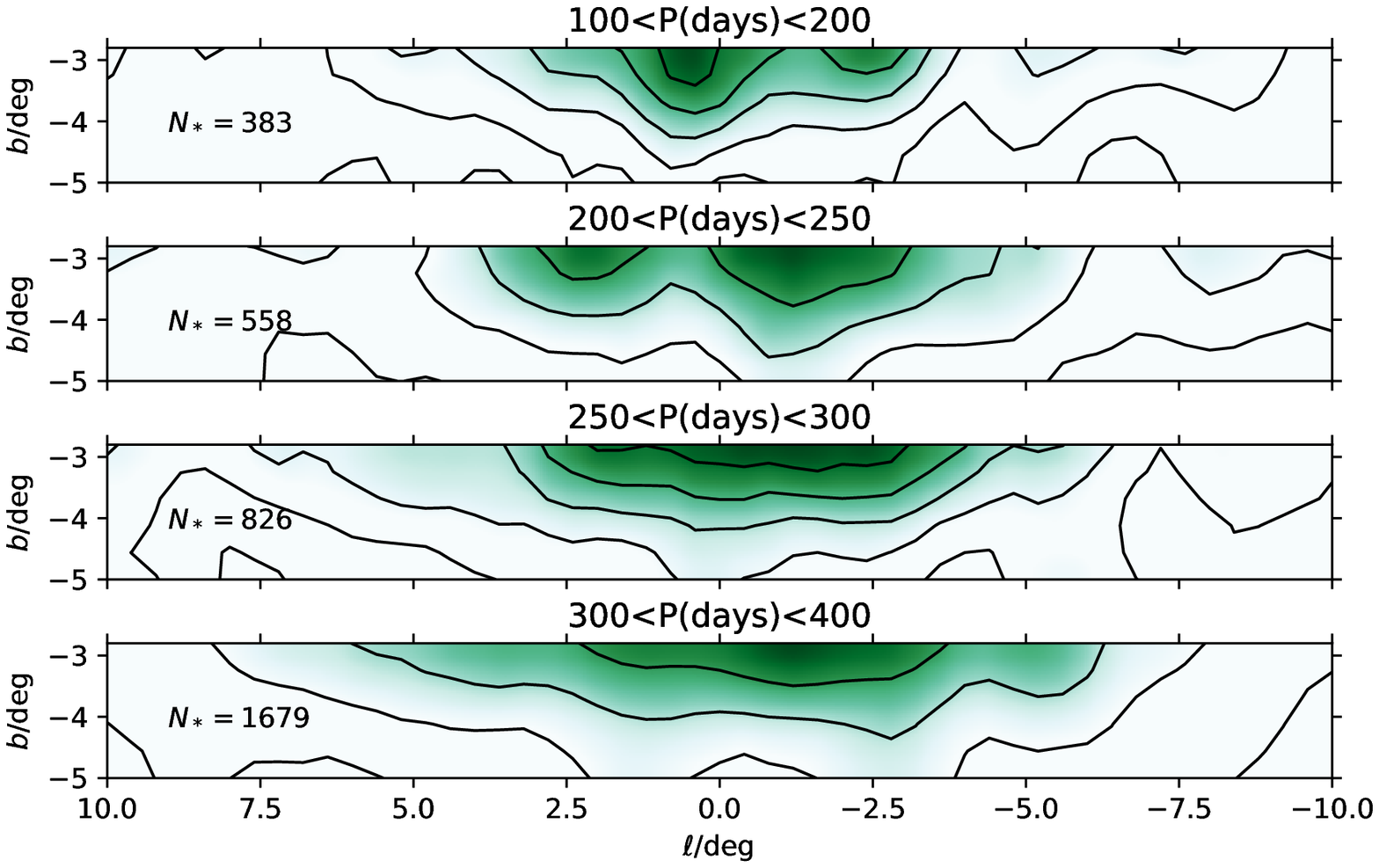}\hfill
    \includegraphics[width=.11\columnwidth]{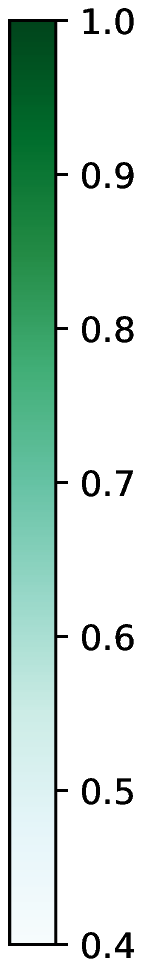}
    \vspace*{-2mm}
    \caption{\textbf{Left:} Number density in galactic coordinates of 4 different stellar populations from the Auriga 6 simulation selected by their age $\tau$. In order to exclude foreground and background contamination we only use stars from within cylindrical radius of 6 kpc. The maps were constructed by a kernel density estimator with smoothing length 1.2\degr. Colour scales are linear and normalized to the respective highest value.
    \textbf{Right:} Density in galactic coordinates of different subsets of Mira stars, selected by their periods. The maps were constructed by a kernel density estimator with smoothing length 1.2\degr. Colour scales are linear and normalized to the respective highest value.
    }
    \label{front}
\end{figure*}

\begin{figure*}
    \centering
    \includegraphics[width=\columnwidth]{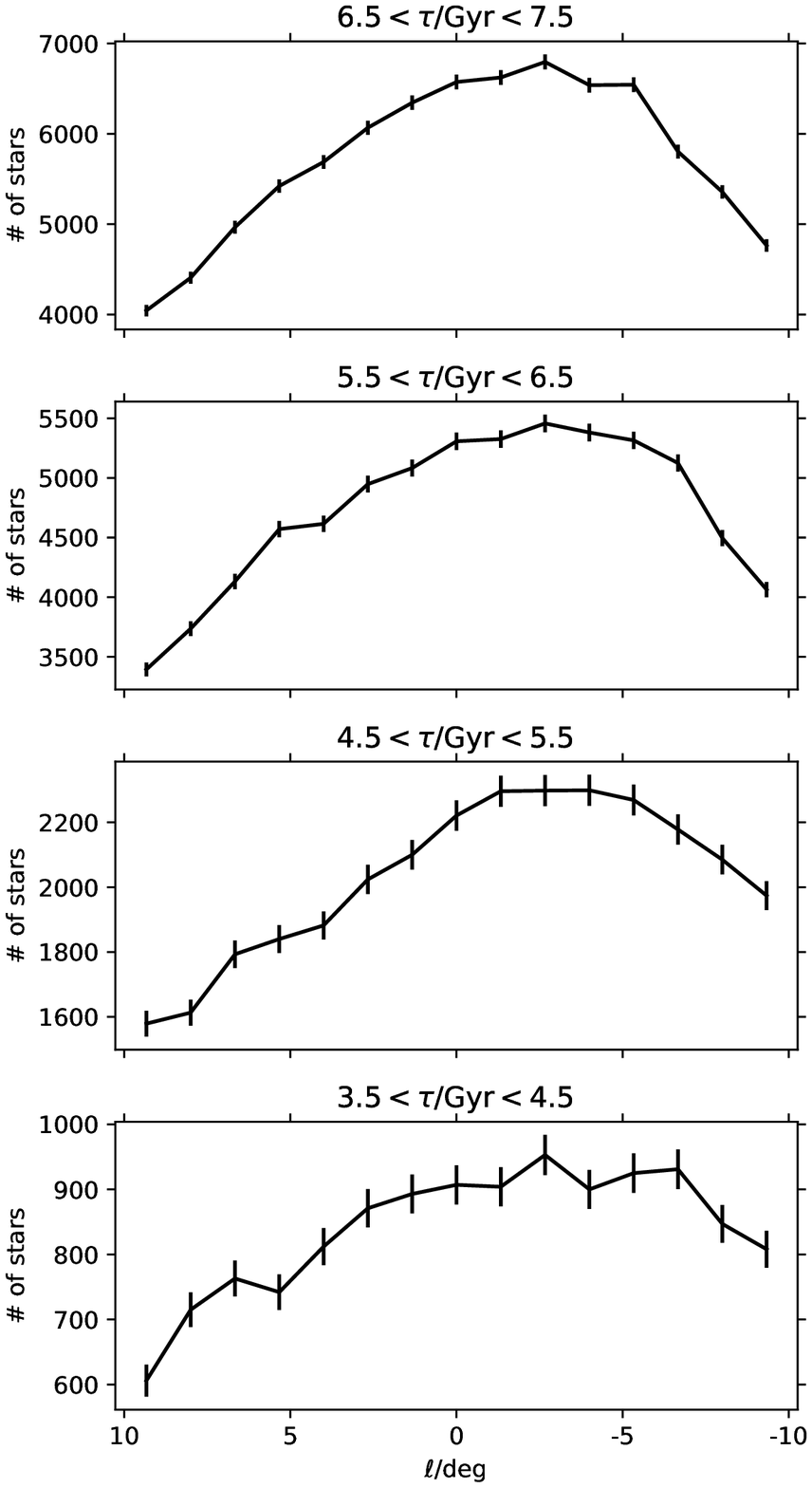}\hfil
    \includegraphics[width=\columnwidth]{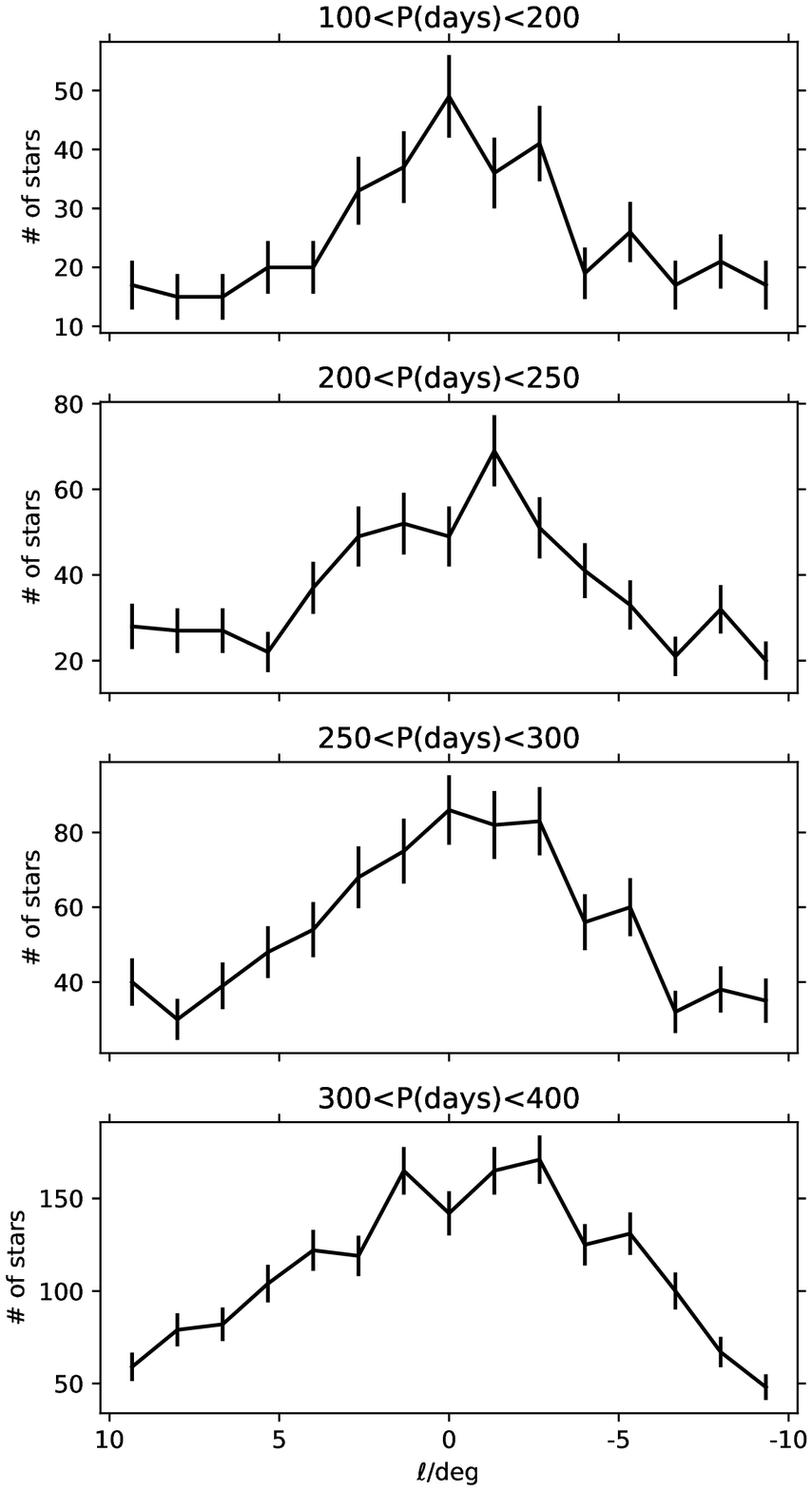}
    \vspace*{-3mm}
    \caption{\textbf{Left:} Star counts as a function of galactic longitude of different subsets of stellar particles from Auriga 6 simulations. \textbf{Right:} Star counts as a function of galactic longitude of different subsets of Mira stars. The selection criteria for the particles and stars are the same as in Figure~\ref{front}.
    }
    \label{hist}
\end{figure*}

In the previous section we showed how a simulation of a barred galaxy with age-morphology dependence for the B/P bulge predicts that the increasing distance between the two humps for younger populations cause the far hump to dominate when survey limitations are present. Here, we use the OGLE Miras catalogue to examine this prediction. It was long known empirically \citep[e.g.][]{Feast2009} and very recently confirmed on theoretical grounds by \cite{Trabucchi2022} that Mira stars obey a period-age relation with longer pulsation periods for younger stars.

Following \cite{Grady2020} we divided the sample described in section 2 into four range of periods 100-200, 200-250, 250-300 and 300-400 days. We plot their smoothed on-sky distribution and histograms in the corresponding region in the right panels of Figures~\ref{front} and~\ref{hist}, respectively. We find that the distributions of Miras in bins with increasing period, corresponding to decreasing age, is consistent with the predictions from simulations, shown in the left panels of the respective figures, that for younger stars the part of the peanut at negative longitude dominates. In general, short-period (older) Miras are more concentrated around $\ell \sim 0\degr$. The over-density of younger populations shifts more towards the negative side (e.g. at $\ell\sim-5\degr$). The older populations (especially 200-250 days) show some weak signs of bi-modality, which is expected, since the small peanut in these populations should fall into the longitude range. However, this bi-modality is not statistically significant due to the low numbers of stars in these bins, unlike the situation for RR Lyrae stars \citep{Semczuk2022}.

The lack of the near hump in younger populations, which in the simulated samples is owed to the longitude cut, may be additionally amplified by the near-hump stars being on average brighter (since closer), which decreases their chance of detection by the OGLE survey.

Comparison of the left with the right panels in Figures~\ref{front} and~\ref{hist} obviously shows that the peanut in the simulation is bigger than what we find from the Mira stars. Indeed, the simulation was picked not to reproduce the Milky Way, but only to show the qualitative effect with ageing populations. We think that the bigger size of the peanut in simulation can be partially attributed to the high disc scale length of the simulation (5.95 kpc), which is roughly twice that of the Milky Way.

\section{Discussion}
\label{sec:discuss}

The period-luminosity relation of Mira stars enabled previous studies of the 3D Milky Way structure, including the B/P bulge. \cite{Lopez2017} combined the \cite{Catchpole2016} and the OGLE-III samples in order to test whether Mira stars follow the X-shape bulge. They concluded that there is no X-shape structure in their sample, which is not surprising considering several factors: their number of stars is low in comparison with the recent OGLE Miras catalogue; additionally, non-negligible distance uncertainties will fill the central dip of the peanut, which is of comparable width to these uncertainties. On top of that, \cite{Lopez2017} combined data from all periods or divided the sample into only two bins. Mixing the populations will lead to a flattening of the expected dip, since the location of the bigger dip of the younger population is at the center, where the older population is hard to resolve (see e.g. Figure~\ref{sim1}). Finally the stars in their work are only analysed at Galactic longitudes not exceeding $|\ell|<8\degr$. For younger populations, where the gap should be easier to find, the location of the close hump, as inferred from our work, is expected somewhere near $\ell\sim10\degr$, since we only find the amplification of the far hump because of the longitude limitation.

More recently \cite*{Grady2020} used Miras from Gaia DR2 \citep{Mowlavi2018} to study the disc and bulge of the Milky Way represented by Miras of different stellar ages. Their conclusion for the B/P bulge is qualitatively consistent with our findings: the peanut morphology is clearer in the youngest population and the bulge is more concentrated in the oldest one. However, a closer look at their Figure 12 compared with our results may suggest some inconsistency. From their smoothed residual map it appears that the far hump at negative longitude is located at $\ell<-9\degr$ and $b<-6\degr$, while our results for the longest period range place the over-density closer to $\ell\sim-3\degr$ and $b\sim-4\degr$. Similarly, they find an over-density on the positive side on the edge of the region considered in our study. It may be that these structures are just an extension of the peanut that is under-sampled in our study and extends in that way to lower latitudes. Alternatively, some over- and under-densities in the on-sky distribution of Gaia DR2 Miras should be affected by the Gaia scanning law covering different regions of the sky with different time coverage and cadence. Miras of longer periods require more time coverage to be classified and this may cause subtle effects, in particular as the boundaries of the Gaia scanning law appear as sharp lines on the sky (see Fig.6 and 7 of \citealt{Mowlavi2018}).

In this study we have shown the expected dependence of the observed B/P bulge morphology on stellar age, in particular a skew density distribution in Galactic longitude, with a peak offset from the Galactic centre longitude. We have demonstrated that the current OGLE Mira sample already shows a trend consistent with these simulations. We conclude that future surveys with more stars will allow for better age splitting and better statistics, and so will shed more light on the age-morphology structure of the Milky-Way bulge. The most imminent improvement will come from Gaia DR3, if the on-sky distribution of variable stars will be largely free from scanning-law effects in relevant regions of the sky. 

\section*{Acknowledgements}
This work was supported by the STFC grant ST/S000453/1. MS thanks B.-E.\ Semczuk for support. EA thanks the CNES for financial support. We appreciate insightful discussions with H.~Aly and P.~Iwanek. RS thanks the Royal Society for generous support via a University Research Fellowship.

\section*{Data Availability}
No data were generated in this study. Any data used are publicly available.



\bibliographystyle{mnras}
\bibliography{example} 





\bsp	
\label{lastpage}
\end{document}